\documentclass[letter]{aa}  

\usepackage{graphicx}
\usepackage{txfonts}
\usepackage{lipsum}
\usepackage{subcaption}         
\usepackage{lscape}             
\usepackage{placeins}

\begin{document}

   \title{Discovery of low-redshift analogues to "Little Red Dots" in DESI: A later evolutionary stage of compact LRDs?}
   \titlerunning{Short title}

   \author{Weiyu Ding\inst{1,} \inst{2,} \inst{3,} \inst{4}, Xu Kong\inst{1,} \inst{3,} \inst{4}\fnmsep\thanks{Corresponding author: xkong@ustc.edu.cn}, Wei-Jian Guo\inst{2}, Hu Zou\inst{2}, \inst{5}\fnmsep\thanks{Corresponding author: zouhu@nao.cas.cn}, Jialai Wang\inst{1,} \inst{3}, Fujia Li\inst{1,} \inst{3,} \inst{4}, Hongxin Zhang\inst{1,} \inst{3}, Jie Song\inst{1,} \inst{3,} \inst{4}, Jingyi Zhang\inst{2}, Niu Li\inst{2}, Wen-Xiong Li\inst{2}  
        }
    \authorrunning{First Author et al.}

   \institute{Department of Astronomy, University of Science and Technology of China, Hefei 230026, P.R. China
            \and National Astronomical Observatories, Chinese Academy of Sciences, Beijing 100101, P.R. China
            \and School of Astronomy and Space Science, University of Science and Technology of China, Hefei 230026, P.R. China
            \and Institute of Deep Space Sciences, Deep Space Exploration Laboratory, Hefei 230026, China
            \and University of Chinese Academy of Sciences, Beijing 100039, People’s Republic of China}

  \abstract 
   {
    The James Webb Space Telescope (JWST) has recently discovered a population of compact, red sources at $z \geq 4$ known as "Little Red Dots" (LRDs). They are characterized by their V-shaped continuum spectra and prominent broad Balmer emission lines. As their underlying physical nature remains debated and direct study at high-redshift is challenging; therefore, we seek to identify and characterize LRD analogues in the low-redshift universe to constrain their properties and potential evolutionary pathways. We identified five candidates at $z = 0.2-0.4$ from the Dark Energy Spectroscopic Instrument (DESI) that exhibit spectral energy distributions (SEDs) and broad Balmer emission lines closely resembling their high-redshift counterparts. However, we find significant differences: our low-redshift sample occupies a different region on the Baldwin, Phillips \& Terlevich (BPT) diagram, and their stellar masses are significantly higher, suggesting a more substantial host galaxy contribution. These sources are not necessarily direct local analogues of high-redshift LRDs, but may represent later evolutionary stages of compact, rapidly accreting systems, or systems with related observational properties arising under different physical conditions. This sample provides a valuable laboratory for detailed follow-up studies to elucidate the nature of LRD-like phenomena.
    }


   \maketitle  \nolinenumbers

\section{Introduction}\label{sec:intro}
Recent JWST observations have uncovered a large population of optically red sources at $z \geq 4$ that appear compact even in JWST imaging, known as LRDs \citep{2023Natur.616..266L, 2023ApJ...954L...4K,2025ApJ...986..126K,2024ApJ...963..129M, 2025ApJ...985..119Z}. These objects display characteristic V-shaped SEDs, defined by a blue rest-frame UV continuum and a red rest-frame optical continuum. Spectroscopic follow-up has shown that a significant fraction ($\sim$ 70–80\%) of LRDs exhibit broad Balmer emission lines with a \rm{FWHM} (Full Width at Half Maximum) $\geq 1000 \, \rm{km \, s^{-1}}$ \citep{2024ApJ...963..129M, 2026ApJ...998..170Z}. This is a key indicator for the presence of actively accreting black holes (BHs) with masses in the range of $10^6 - 10^8 M_{\odot}$ \citep{2025ApJ...978...92L, 2025ApJ...986..126K}. The physical nature of LRDs remains a subject of ongoing debate, with several theoretical models proposed to explain their characteristic V-shaped SED \citep{2025ApJ...988L..22I,2026ApJ...997..364L}. Explanations for this distinctive UV-optical turnover generally fall into two categories. The first hypothesis posits a non-AGN origin, attributing the continuum to a massive and compact stellar population \citep{2024ApJ...968....4P, 2024ApJ...977L..13B, 2024ApJ...969L..13W, 2025ApJ...981..191M}. The second, more prevalent, category of models invokes an active galactic nucleus (AGN). In these scenarios, the SED is attributed to an accretion disk that is either: (i) heavily obscured by gas \citep{2025ApJ...980L..27I,2025arXiv250316596N,2025ApJ...994..113L}; (ii) viewed through a non-standard dust attenuation curve \citep{2025ApJ...980...36L} or (iii) intrinsically extended and gravitationally unstable \citep{2026NatAs.tmp...41Z}.

Interpreting the UV–optical turnover as stellar implies high stellar masses ($> 10^{9-10} M_{\odot}$) for LRDs, derived primarily from the strong Balmer breaks in this region. These large masses, combined with the high comoving-volume density of LRDs, yield a stellar-mass function potentially inconsistent with the standard $\Lambda$-CDM cosmological model \citep{2024ApJ...973L..49I}. Furthermore, these high stellar masses are incompatible with the low dynamical masses inferred from their emission line widths. It has recently been proposed that this tension can be alleviated if the observed continua, including the strong Balmer breaks, are dominated by AGN accretion discs rather than stellar populations \citep{2025ApJ...980L..27I}. In this scenario, the Balmer breaks are attributed to dense nuclear gas, potentially close to the broad-line region (BLR), that is optically thick to the Balmer continuum, thus avoiding the need for massive stellar populations. Recent models also explore the origin of the obscuring gas, including "quasi-star" models in which a direct-collapse black hole retains a gaseous envelope \citep{2026ApJ...996...48B, 2025arXiv250316596N}, or turbulent accretion flows surrounding the black hole \citep{2025ApJ...994..113L}. However, the AGN interpretation is not without challenges. Notably, LRDs exhibit peculiar properties not common in local AGN or bright quasars: extremely weak X-ray emission, the widespread absence of high-ionization emission lines and hot-dust torus emission, weak or no continuum variability, and $\sim$20\% of them show strong, non-stellar Balmer-line (and sometimes helium-line) absorption \citep{2025ApJ...995...24K, 2024ApJ...963..129M, 2025MNRAS.538.1921M, 2025ApJ...985..119Z}. A full investigation of the AGN scenario is often limited by the lack of constraining multi-wavelength data at high redshift \citep{2025A&A...701A.168D} and of multi-epoch observations \citep{2025ApJ...995...24K}.

The immense distances of JWST LRDs present a significant obstacle to detailed physical characterization. Key properties of their host galaxies, stellar populations, and gas kinematics remain largely unresolved, making it difficult to definitively distinguish between the competing theoretical models. This observational challenge highlights the crucial need to find local analogues \citep{2026ApJ...997..364L, 2026MNRAS.545f2235J}. Despite the pronounced cosmic evolution of LRDs, with their number density decreasing by 1–2 dex from $z\sim4$ to $2$ \citep{2025ApJ...986..126K, 2026ApJ..1000...59M}, local counterparts remain exceedingly rare, yet they represent highly valuable targets. These nearby laboratories provide a unique opportunity to probe the physical conditions that give rise to the LRD phenomenon with a fidelity impossible at high redshift, offering critical constraints on the evolutionary pathways of LRDs and their connection to the cosmic growth of SMBHs. In this Letter, we perform a systematic search for low-redshift LRD analogues using the dataset from DESI \citep{2016arXiv161100036D}. In Section \ref{sec:data}, our selection yields five objects in the redshift range of $z = 0.2-0.4$ that reproduce the defining characteristics of the LRD. We present our analysis of these sources and discuss their implications in Section \ref{sec:results}, followed by a summary in Section \ref{sec:summary}.

\section{Data and sample} \label{sec:data}

We identify five LRD analogues at $z = 0.2-0.4$ from the DESI Data Release 1 \citep[DR1;][]{2025arXiv250314745D}. Candidates were selected primarily based on their V-shaped SEDs and the specific selection method is detailed in Appendix \ref{sec:appendixA}. We supplement the DESI spectra with archival photometric measurements covering wavelengths from the optical to the mid-infrared (MIR) to construct multi-wavelength SEDs. We note that our selection algorithm does not impose explicit constraints on emission line widths, a key feature of the high-redshift LRD population. Nonetheless, all five sources satisfying these criteria exhibit clear broad H$\alpha$ emission in their DESI spectra. 

   \begin{figure}[htbp]
   \centering
   \includegraphics[width=0.9\hsize]{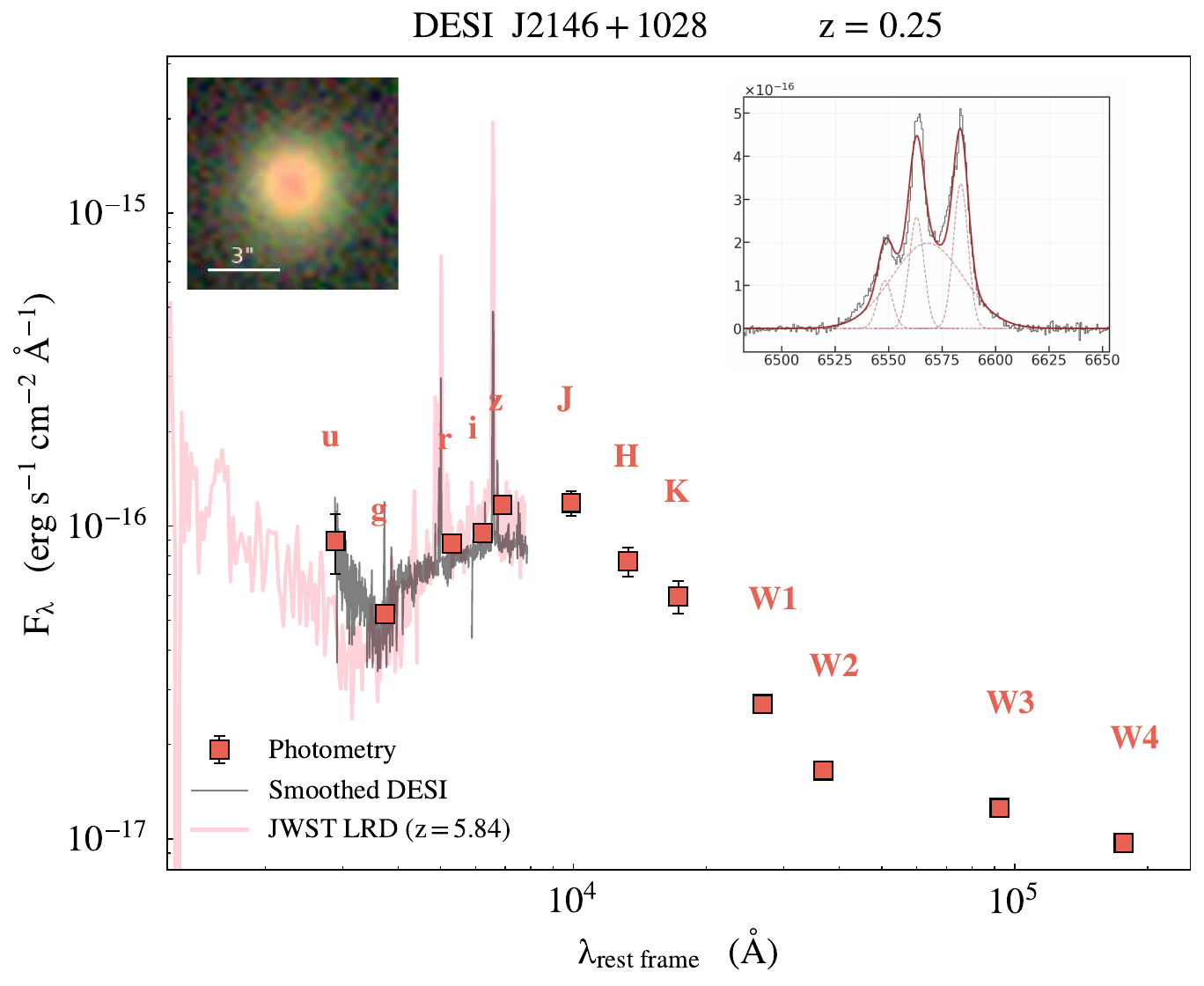}
      \caption{Multi-wavelength view of an LRD analogue at $z=0.25$. Main panel: Rest-frame SED showing archival photometry (red squares) and the smoothed DESI spectrum (black line). For comparison, the rest-frame JWST/NIRSpec PRISM spectrum of a high-redshift LRD is over-plotted in pink. Top-left inset: DESI Legacy Imaging Surveys $grz$-composite image \citep{2019AJ....157..168D} showing the compact morphology. Top-right inset: Zoomed-in rest-frame DESI spectrum and fitting models of the H$\alpha$ region.
      }
         \label{fig:sed}
   \end{figure}

Fig. \ref{fig:sed} presents the multi-wavelength data—including imaging, photometry, and the DESI spectrum—for a representative low-redshift analog DESI J2146+1028. We over-plot a template JWST/NIRSpec PRISM spectrum of a LRD at $z=5.84$ \citep{2025ApJ...995..118S} to highlight the strong similarity between the two populations. The pronounced V-shaped continuum, broad H$\alpha$ emission and the declining mid-infrared slope seen in this example are characteristic of all five sources in our sample. The data for the remaining four analogues are shown in Fig. \ref{fig:sed_figures}.

\section{Results and discussion}\label{sec:results}
In this section, we perform a detailed analysis of our five LRD analogues. Using their DESI spectra, we investigate the ionization state of their interstellar gas via emission-line diagnostics and estimate their black hole masses. We then discuss the potential origins of the observed differences between these low-redshift analogues and the high-redshift LRD population.

\subsection{BPT Diagram} \label{subsec:BPT}
   \begin{figure*}[htbp]
   \centering
   \includegraphics[width=0.9\hsize]{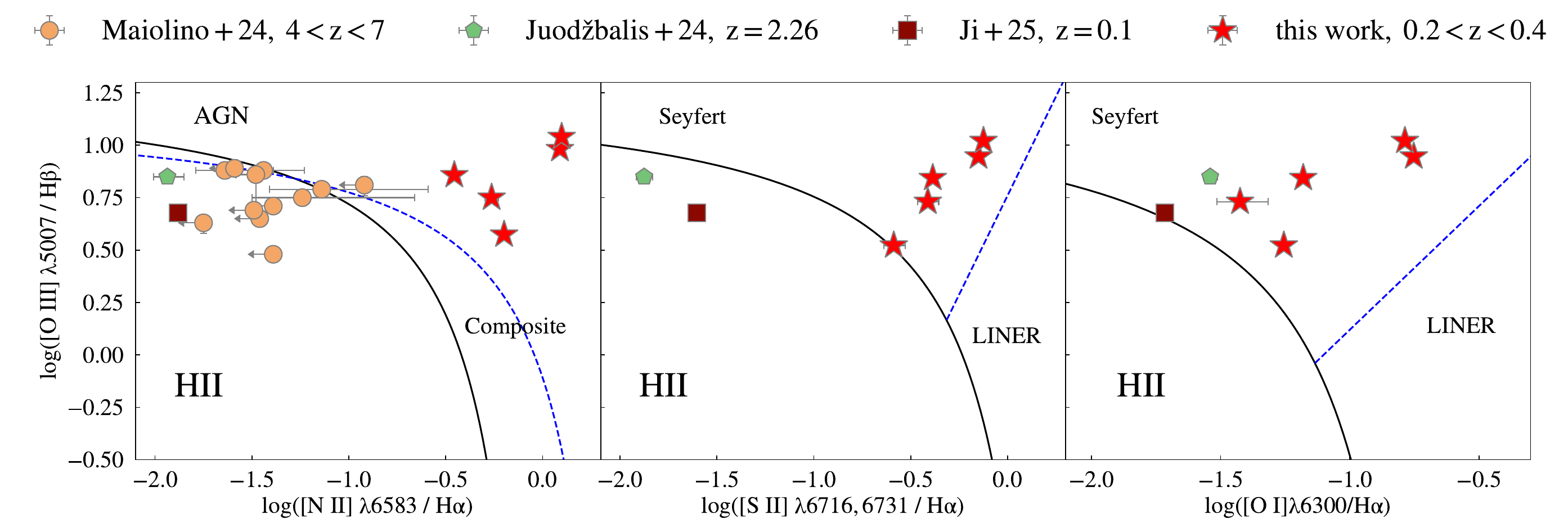}
      \caption{Locations of five LRD analogues in BPT/VO diagrams \citep{1981PASP...93....5B, 1987ApJS...63..295V} with black solid and blue dashed demarcation lines from \cite{2001ApJ...556..121K, 2006MNRAS.372..961K, 2003MNRAS.346.1055K}. Our sample are shown in red stars. For comparison, we show high-redshift LRDs from \cite{2024ApJ...963..129M} as orange circles. We also show the “Rosetta Stone” at $z=2.26$ \citep[green pentagon;][]{2024MNRAS.535..853J} and a previously identified local LRD at $z=0.1$ \citep[brown square;][]{2026MNRAS.545f2235J}.}
         \label{fig:BPT}
   \end{figure*}

We examine the narrow emission-line excitation diagnostics for our sample. Fig. \ref{fig:BPT} shows the positions of the five LRD analogues on the [N {\sc ii}], [S {\sc ii}], and [O {\sc i}] BPT/VO diagnostic diagrams \citep{1981PASP...93....5B, 1987ApJS...63..295V}. For comparison, we also plot high-redshift LRDs at $4<z<7$ selected by \cite{2024ApJ...963..129M}, the closest LRD observed by JWST at $z=2.26$ \citep[the “Rosetta Stone”;][]{2024MNRAS.535..853J} and a previously identified local LRD at $z=0.1$ \citep{2026MNRAS.545f2235J}. 

In the [N {\sc ii}]- and [S {\sc ii}]-based diagrams, all five of our LRD analogues lie unambiguously in the AGN region, well above the theoretical maximum starburst line of \cite{2001ApJ...556..121K}. This placement contrasts sharply with both the high-redshift LRD population and other known low-redshift analogues, which typically reside near the demarcation line separating H {\sc ii} regions from AGN. While the [O {\sc iii}]/H$\beta$ ratios of our sample are consistent with those of previously studied LRDs, their [N {\sc ii}]/H$\alpha$ and [S {\sc ii}]/H$\alpha$ ratios are significantly elevated. This offset is an indicator of a metallicity effect. The characteristically low [N {\sc ii}]/H$\alpha$ and [S {\sc ii}]/H$\alpha$ ratios in high-redshift LRD populations are commonly attributed to the low gas-phase metallicity of those systems \citep{2024A&A...691A.145M, 2025A&A...697A.175S, 2026MNRAS.546ag086J}. The position of our sample therefore implies a more chemically evolved, higher-metallicity environment. Using the N2-based method described in \cite{2004MNRAS.348L..59P}, we calculate an average metallicity for our sample of $\sim 1.24~Z_{\odot}$. This is significantly higher than that found for high-redshift LRDs \citep[typically $\rm{Z\sim0.2 \,Z_{\odot}}$; ][]{2024A&A...691A.145M}.

The [O {\sc i}]-based VO diagram provides a more definitive classification, as the [O {\sc i}]$\lambda$6300 emission line is less sensitive to metallicity than the [N {\sc ii}]$\lambda$6583 and [S {\sc ii}]$\lambda \lambda$6716,6731 lines \citep{2020MNRAS.496.1262J}. In this diagnostic, our sample and the other low-redshift LRDs occupy the Seyfert region. The clean separation in the [O {\sc i}]-based VO diagram thus confirms an AGN as the dominant ionization source and suggests that it is a more robust classifier for LRDs across cosmic time.

\subsection{$M_{\rm{BH}}$–$M_*$ Relation} \label{subsec:mass}

   \begin{figure}[htbp]
   \centering
   \includegraphics[width=0.9\hsize]{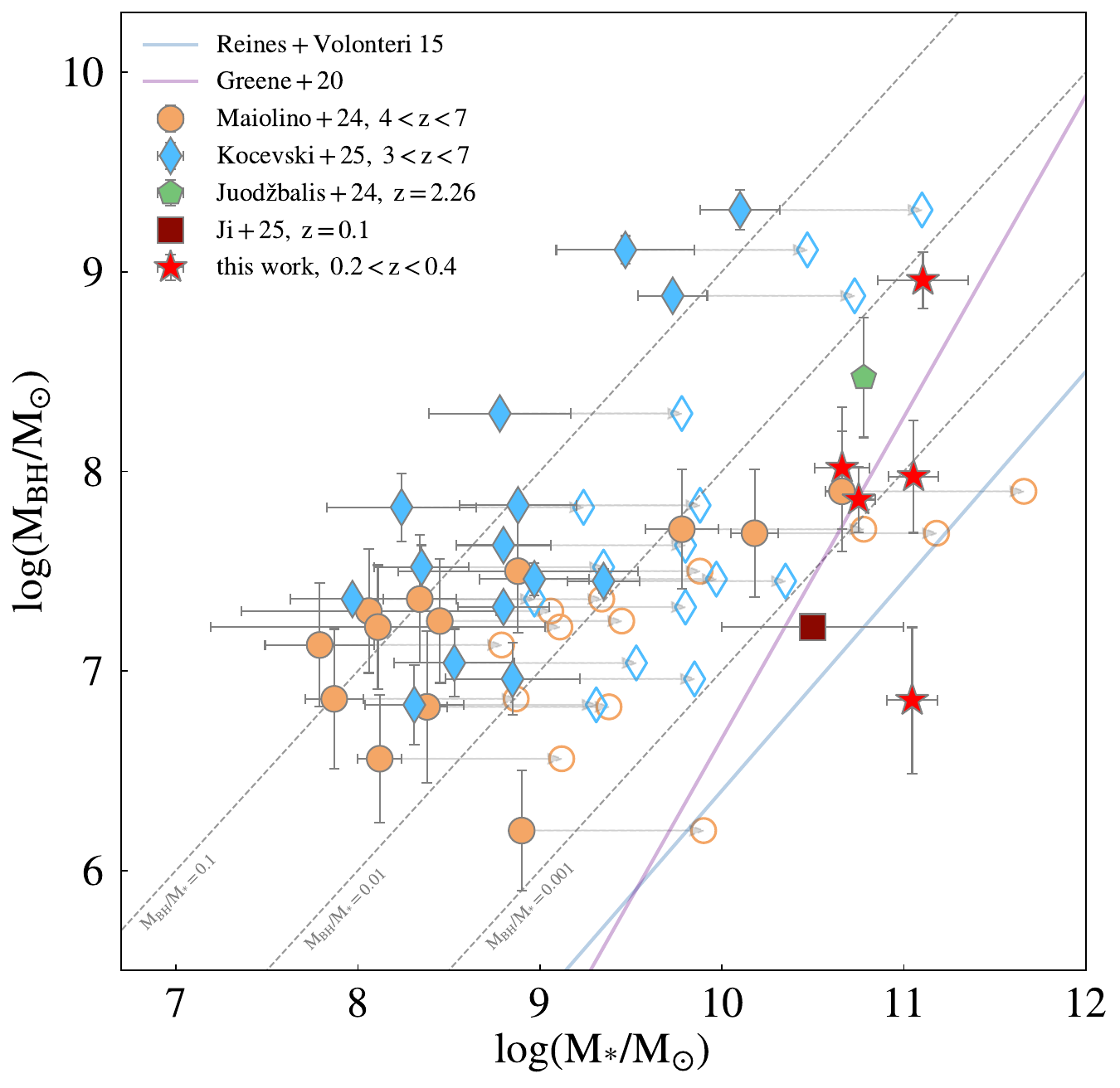}
      \caption{Relation between black hole mass and the host galaxy stellar mass. Our low-redshift LRD analogues are shown as red stars. For comparison, we also include high-redshift LRDs from JWST surveys and other low-redshift LRDs. The established local scaling relations for active galaxies from \cite{2015ApJ...813...82R} and inactive galaxies from \cite{2020ARA&A..58..257G} are shown as the colored dashed lines respectively. Black dashed lines represent different ratios of black hole mass to total stellar mass ($M_{BH}/M_{\star}$ = 0.1, 0.01 and 0.001). The arrows and hollow markers show the effect of increasing the stellar masses of the high-redshift LRDs by a factor of 10.}
         \label{fig:MbhMste}
   \end{figure}

We investigated the physical properties of five LRD analogues, estimating their stellar masses ($M_{\star}$) through multi-wavelength SED fitting. Our decomposition reveals a dichotomy in the sample: three sources are best fit by a heavily obscured AGN, while the remaining two are dominated by an unobscured AGN component. The details of the SED fitting are shown in Appendix \ref{sec:appendixB}. We estimate black hole masses ($M_{\rm{BH}}$) using virial scaling relations, though we acknowledge ongoing debate regarding their applicability to LRDs. Notably, \cite{2026Natur.649..574R} suggests that scatter-broadening in dense, ionized environments may cause systematic overestimation of virial masses in these sources. Consequently, our $M_{\rm{BH}}$ values should be treated with caution and potentially viewed as upper limits. Nevertheless, to ensure consistency with existing high-redshift LRD literature and facilitate comparative analysis, we retain this standard virial approach. For this work, we adopt the specific calibration for the H$\alpha$ line from \cite{2015ApJ...813...82R} and the derived black hole masses for our sample ranging from $7.14 \times 10^{6}$ $M_\odot$ to $9.09 \times 10^8$ $M_\odot$. We show the observed and derived properties of the LRD analogues in Appendix \ref{sec:appendixC}.

Fig. \ref{fig:MbhMste} shows the black-hole mass of LRD analogues on the $M_{BH}$ versus $M_{*}$ plane. While high-redshift LRDs \citep{2024A&A...691A.145M, 2025ApJ...986..126K} are typically over-massive relative to their host stellar masses compared with local scaling relations \citep{2015ApJ...813...82R, 2020ARA&A..58..257G}, our sample and other low-redshift LRDs \citep{2024MNRAS.535..853J, 2026MNRAS.545f2235J} are consistent with these local relations. Stellar masses for the high-redshift LRDs are subject to large model uncertainties \citep{2024ApJ...977L..13B}. However, even if we adopt stellar masses $10\times$ larger (as indicated by the arrows and hollow markers in Fig. \ref{fig:MbhMste}), they would still be over-massive relative to their host galaxy stellar mass. This difference arises because, although the black hole masses of our sample are comparable to their high-redshift counterparts, their host galaxy stellar masses are $\sim 2-3$ dex higher. 

\subsection{An Evolutionary Link or Two Distinct Populations} \label{subsec:discussion}

Our analysis identifies low-redshift objects that mimic the spectral hallmarks of high-redshift LRDs: a V-shaped continuum, broad Balmer lines and compact imaging. However, we find fundamental physical discrepancies between the populations. However, unlike many high-redshift LRDs, our sample shows only H$\alpha$ emission lines and no H$\alpha$ absorption features. While high-redshift LRDs are metal-poor and host "over-massive" black holes relative to their host galaxies, our sample reside in chemically evolved and massive hosts. Despite comparable black hole masses, the stellar masses of our sample are $\sim 2$ dex higher than their high-redshift counterparts. Furthermore, our sample exhibits two distinct SED configurations, both of which deviate from the "standard" LRD interpretative framework.

These findings suggest two primary interpretations. First, these systems may represent a late-stage evolutionary phase of the LRD population. Within this framework, the observed SED diversity could stem from different physical LRD models, ranging from stellar-dominated UV/optical model or the unobscured AGN-dominated models (see Appendix \ref{sec:appendixB}). Alternatively, these sources might reflect various stages of a dispersing gaseous cocoon. In this scenario, a super-Eddington accreting SMBH originally enshrouded in dense gas undergoes dispersal \citep{2025arXiv251202096F}; the obscured sources indicate a transition to a porous structure, while the unobscured sources mark a nearly dissipated state. Regardless of the specific LRD model or dispersal stage, this evolutionary pathway implies that the host galaxy undergoes substantial growth and chemical enrichment over cosmic time. While high-redshift LRDs are characterized by high $M_{ BH}/M_{\ast}$ ratios in metal-poor environments \citep{2025ApJ...983...60C}, mass assembly from $z \gtrsim 4$ to $z \sim 0.3$ builds up the stellar component. Our sample would thus represent LRD-like activity that is either more evolved or perhaps reactivated within more massive, metal-rich hosts.

Second, our sample and high-redshift LRDs may be intrinsically distinct. The resemblance between their V-shaped SEDs might be phenomenological rather than physical, with high-redshift signatures originating from conditions unique to the early Universe, while low-redshift analogues are driven by distinct mechanisms in mature environments. Discriminating between these scenarios is essential. An evolutionary link would provide a unique laboratory to study late-stage LRD transitions, whereas unrelated populations would imply the V-shaped SED is a degenerate signature of diverse pathways. Spatially resolved spectroscopy will be critical to deblend stellar and AGN emission to test these models.

\section{Summary}\label{sec:summary}

We have conducted a systematic search for low-redshift analogues to the high-redshift LRD population within the DESI DR1 spectroscopic survey. Our search identified five objects at $0.2 < z < 0.4$ that share the defining characteristics of the LRD. Our findings are summarized below.

\begin{itemize}
    \item Unlike high-redshift LRDs, our low-redshift analogues are unambiguously classified as AGN by BPT/VO diagnostics. This suggests they reside in more chemically mature, relatively metal-rich host galaxies.
    \item The black hole masses of our sample ($7.14 \times 10^{6}$ $M_\odot$ to $9.09 \times 10^8$ $M_\odot$) are comparable to their high-redshift counterparts. However, their host galaxy stellar masses are significantly higher by 2 orders of magnitude.
    \item Consequently, our low-redshift analogues are consistent with the local $M_{\mathrm{BH}} - M_{\star}$ relation, in contrast to high-redshift LRDs.
\end{itemize}

These properties suggest two compelling possibilities. We propose that our low-redshift analogues may represent a more mature phase of the LRD phenomenon, they are evolutionary descendants of the high-redshift population. Alternatively, these objects may be an intrinsically distinct population, with their V-shaped SEDs arising from a different physical mechanism. Future spectroscopic observations will be crucial in mapping the kinematics and stellar populations of these unique systems, providing a definitive test of their nature.

\begin{acknowledgements}

This work is supported by the National Science Foundation of China (NSFC, Grant Nos. 12233008, 12573012, 12120101003, 12373010, 12173051), the National Key R\&D Program of China (Grant Nos. 2023YFA1608100, 2023YFA1607804, 2022YFA1602902, 2023YFA1607800), the Strategic Priority Research Program of the Chinese Academy of Sciences (Grant Nos. XDB0550200, XDB0550100, XDB0550000), the Cyrus Chun Ying Tang Foundations, the 111 Project for "Observational and Theoretical Research on Dark Matter and Dark Energy" (B23042). We also acknowledge support from the China Manned Space Project (Grant No. CMS-CSST-2025-A06) and the National Astronomical Observatories of the Chinese Academy of Sciences (Grant Nos. E5ZQ7801, E5ZB7801, E4TG2001).

\end{acknowledgements}

\bibliographystyle{aa}
\bibliography{references}

\begin{appendix}
\onecolumn
\section{Finding LRD analogues from DESI}\label{sec:appendixA}

We utilize the emission-line flux measurements provided by the public \textit{FastSpecFit} \footnote{ \url{https://data.desi.lbl.gov/doc/releases/dr1/vac/fastspecfit/}} pipeline to search for low-redshift LRD analogues. Line fluxes are measured via Gaussian profile fitting. Due to their proximity, the following line groups are defined: Mg {\sc ii}$\lambda$$\lambda$ 3727,3729; [O {\sc ii}]$\lambda$$\lambda$ 3727,3729; H$\gamma$ and [O {\sc iii}]$\lambda$4363; [O {\sc iii}]$\lambda$$\lambda$ 4959,5007;[O {\sc i}]$\lambda$6300 and [S {\sc iii}]$\lambda$6312; [N {\sc ii}]$\lambda$$\lambda$ 6548,6583 and H$\alpha$; [S {\sc ii}]$\lambda$$\lambda$ 6716,6731; [Ar {\sc iii}]$\lambda$7135 and [O {\sc ii}]$\lambda$$\lambda$ 7320,7330. Emission-line fluxes in each group are fitted with Gaussian functions simultaneously in order to mitigate the contamination between adjacent features. The background-subtracted emission-line profiles are modeled with Gaussian functions, where the mean, standard deviation, and amplitude are free parameters. Uncertainties are derived from the fit using the error spectrum as a weight. For lines exhibiting the $\mathrm{FWHM > 1000 \, km \, s^{-1}}$, we perform a multi-component fit consisting of narrow and broad Gaussian profiles. A broad component is formally identified if its velocity dispersion exceeds that of the narrow component and the resulting $\chi^2$ significantly improves the fit.

Our parent sample consists of the 17,995,820 unique targets within this catalog. From this sample, we first select extragalactic objects by requiring SPEC-TYPE $\neq$ "STAR". We then apply a series of quality cuts to ensure robust redshift determinations, selecting only objects with a redshift warning flag ZWARN = 0 and a minimum redshift confidence of $\Delta \chi^2 > 15$ \citep{2023AJ....165...58Z}. 

We impose a cut on the rest-frame EW of the [O {\sc iii}]$\lambda$5007 emission line, requiring EW([O {\sc iii}]) $\geq$ 10 {\AA}. This criterion effectively removes stellar contaminants, quiescent galaxies, and post-starburst systems, which are prevalent at low-redshift and whose strong Balmer breaks could otherwise mimic the desired spectral shape \citep{2026ApJ...997..364L}. Subsequently, to select for the compact morphologies characteristic of LRDs, we require an $r$-band half-light radius of less than 1 arc second, as derived from the DESI pipeline's galaxy profile fit. 

   \begin{figure*}[htbp!]
        \centering
        \includegraphics[width=0.45\linewidth]{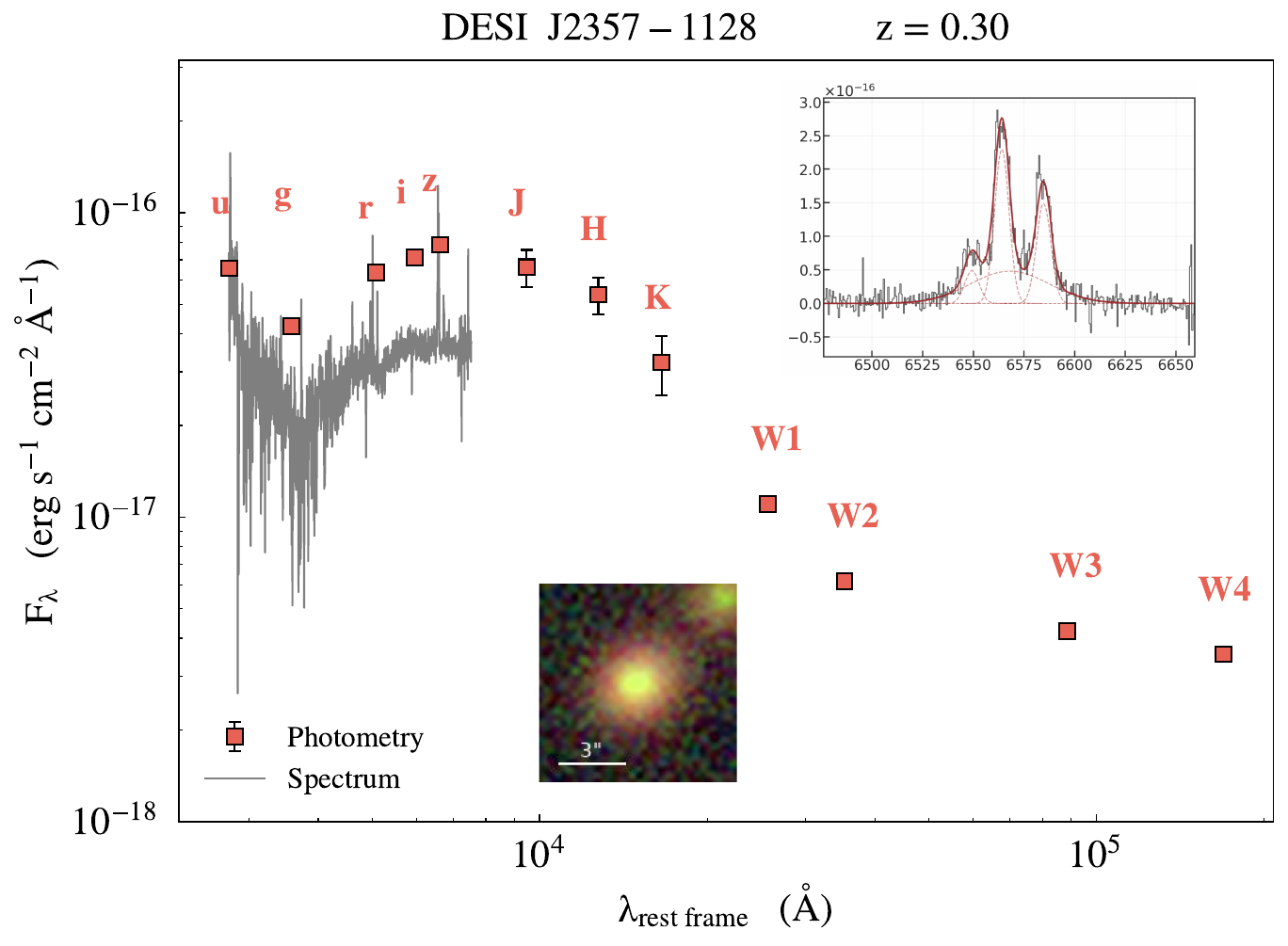}
        \includegraphics[width=0.45\linewidth]{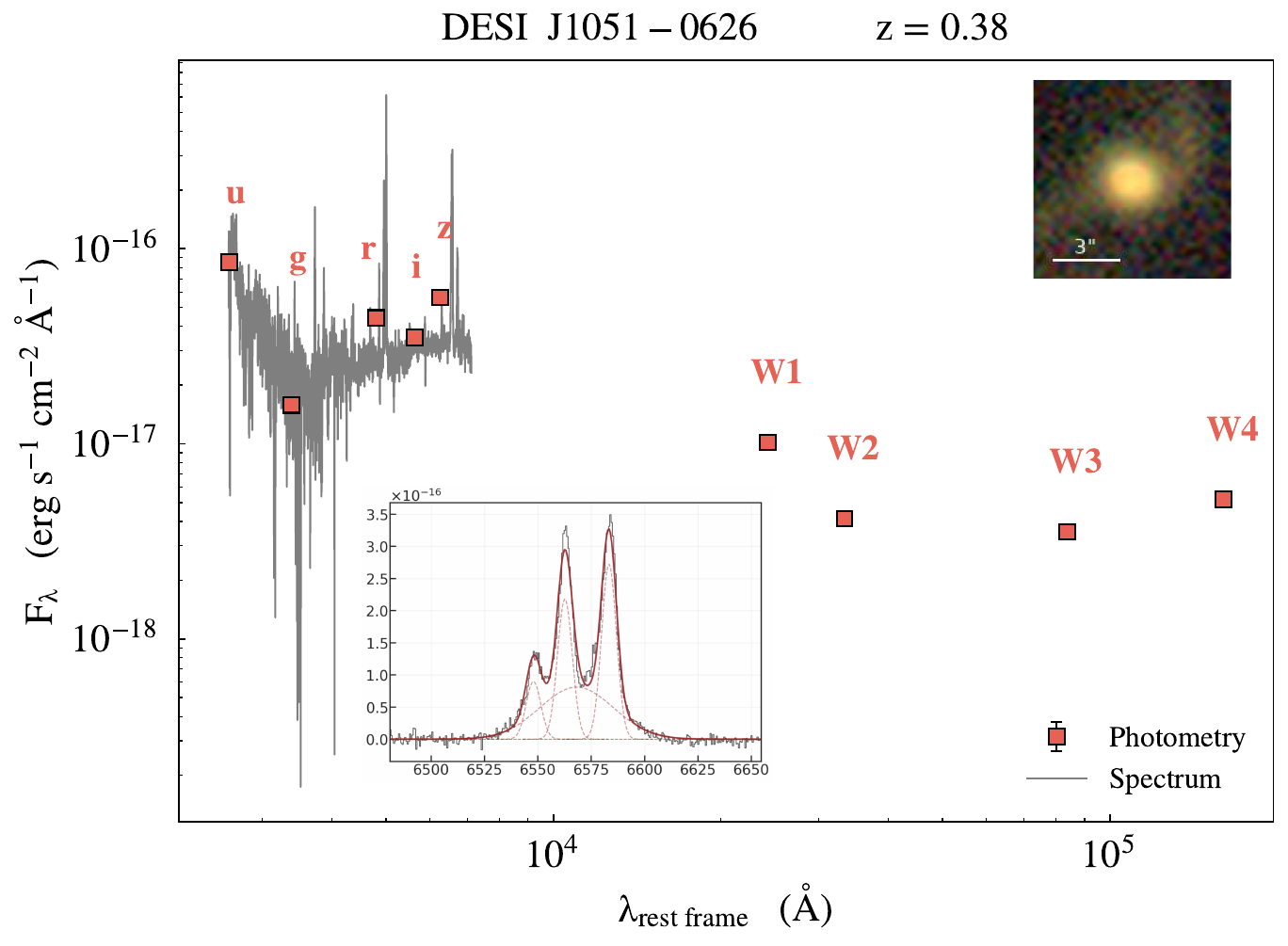}
        \includegraphics[width=0.45\linewidth]{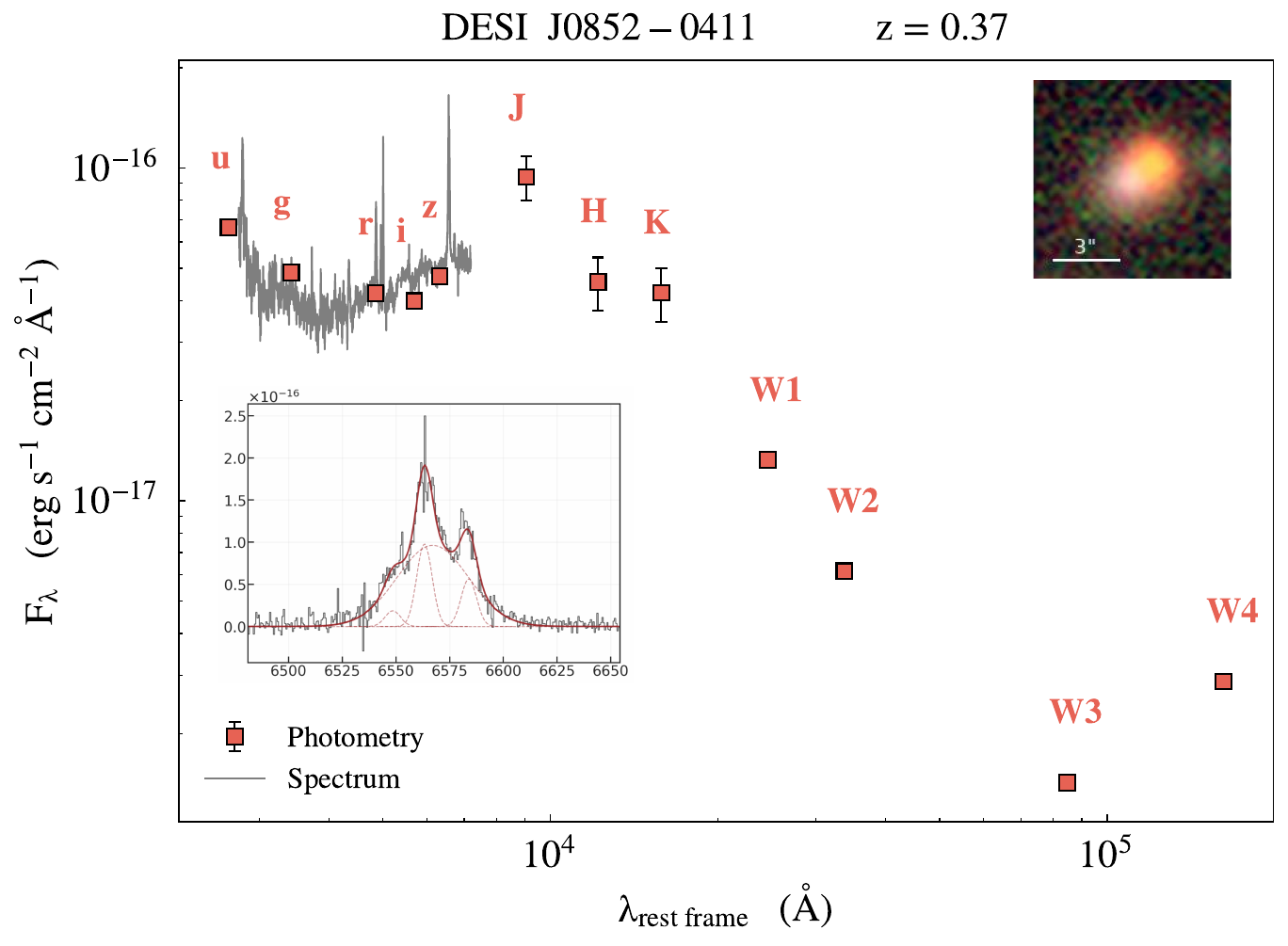}
        \includegraphics[width=0.45\linewidth]{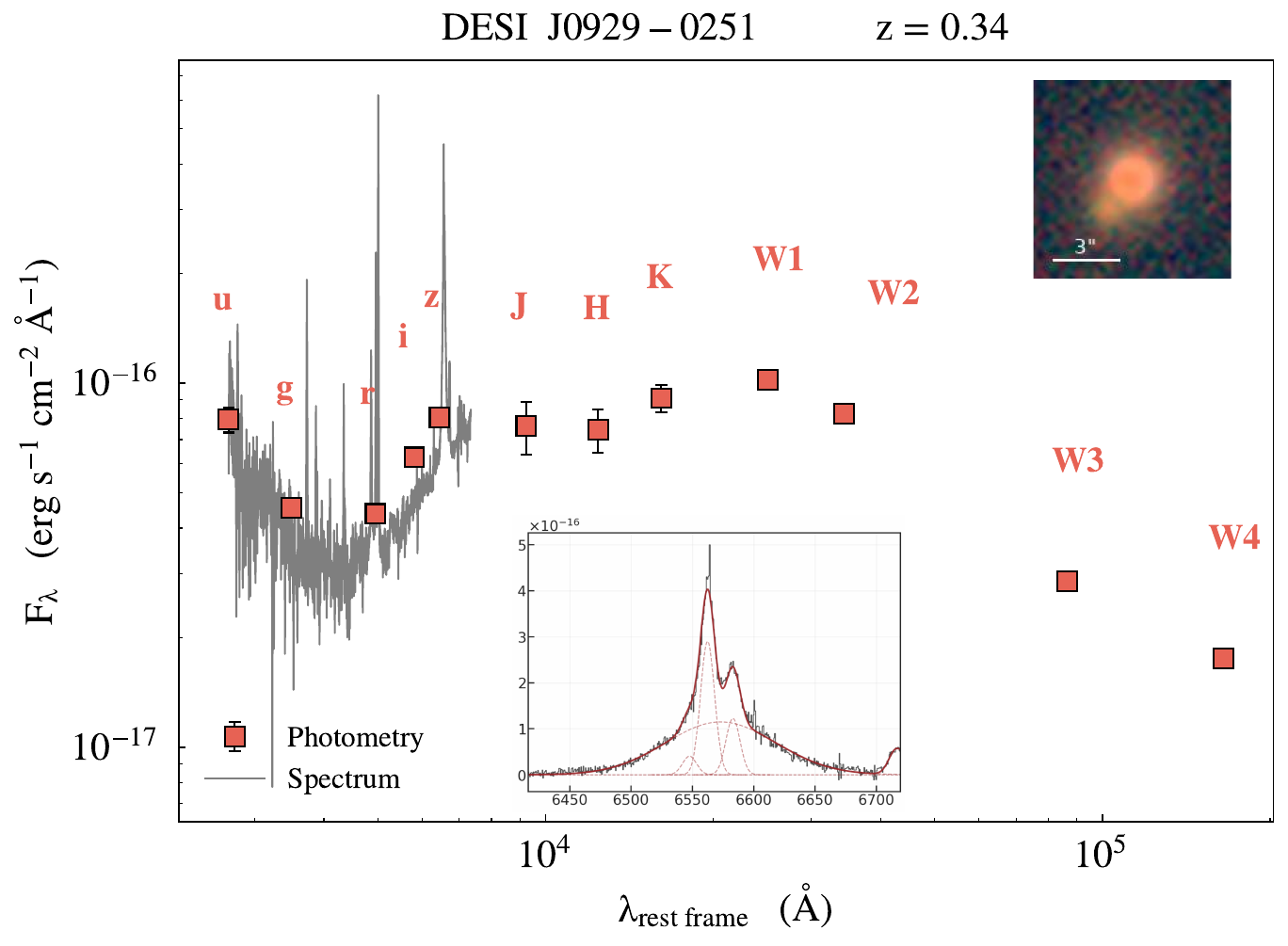}
        \caption{Multi-wavelength data for the other four LRD analogues in our final sample. Each panel includes the DESI Legacy Surveys $grz$-composite image, archival photometry, and the DESI spectrum.}
        \label{fig:sed_figures}
    \end{figure*}

We apply our selection algorithm to the rest-frame DESI spectra to identify V-shaped continua. The search is confined to a redshift of $z \leq 0.5$  which ensures the H$\alpha$ emission line remains within the DESI spectral coverage (3600–9800 {\AA}). To ensure data quality, we additionally require a median continuum signal-to-noise ratio (S/N) greater than 5. In order to isolate the continuum, all significant emission lines in each spectrum are masked. We then model the continuum as a power law, $F_{\lambda} \propto \lambda^{\beta}$, and perform separate fits for the rest-frame ultraviolet (UV) and optical regimes to determine their respective slopes, $\beta_{UV}$ and $\beta_{opt}$. These regimes are defined by wavelengths blue-ward ($\lambda_{\rm{rest}} \leq 3645${\AA}) and red-ward ($\lambda_{\rm{rest}} \geq 3645${\AA}) of the Balmer break. Following the methodology of previous LRD studies \citep{2025ApJ...986..126K, 2026ApJ...998..170Z}, we select LRD candidates that satisfy the V-shaped criteria: a red optical slope $\beta_{\rm{opt}} > 0$ and a blue UV slope $\beta_{\rm{UV}} < -0.37$.

Our full selection procedure yields a final sample of seven LRD candidates. To construct the multi-wavelength SED for our candidates, we compiled archival photometric data spanning from the optical to the MIR. This dataset includes \textit{u}-band photometry from the Sloan Digital Sky Survey \citep[SDSS;][]{2009ApJS..182..543A}; \textit{g, r, i, z}-bands from the DESI Legacy Imaging Surveys \citep[LS;][]{2019AJ....157..168D} DR10; near-infrared J, H, and K-bands from the The Two Micron All Sky Survey \citep[2MASS;][]{2006AJ....131.1163S}; and W1, W2, W3, and W4-bands from the Wide-field Infrared Survey Explorer \citep[WISE;][]{2010AJ....140.1868W}. We used these SEDs to perform a final sample vetting, which led to the exclusion of two of the seven objects. We removed the first source because of a significant inconsistency between its spectroscopic and photometric data. The second was rejected because its infrared continuum lacks the characteristic downturn expected from AGN torus emission.  We also searched for counterparts in the GALEX \citep{2005ApJ...619L...1M} ultraviolet and Chandra \citep{2024ApJS..274...22E} X-ray archives but found no matches for our final candidates. We note that many of our sources do not have corresponding SDSS $u$-band measurements, either because they lie outside the SDSS footprint or lack reliable detections. To ensure a consistent comparison across the sample, we therefore construct synthetic photometry by convolving the DESI spectra with the SDSS $u$-band filter transmission curve. We further note that the DESI spectral coverage does not fully extend to the SDSS $u$-band wavelength range. The $u$-band flux is therefore estimated by extrapolating the observed spectrum using a power-law function. It shoud be noted that synthetic photometry may involve additional sources of uncertainty. However, we emphasize that these synthetic $u$-band points are used only for visualization and illustrative purposes.

 \begin{figure*}[h!]
        \centering
        \includegraphics[width=0.45\linewidth]{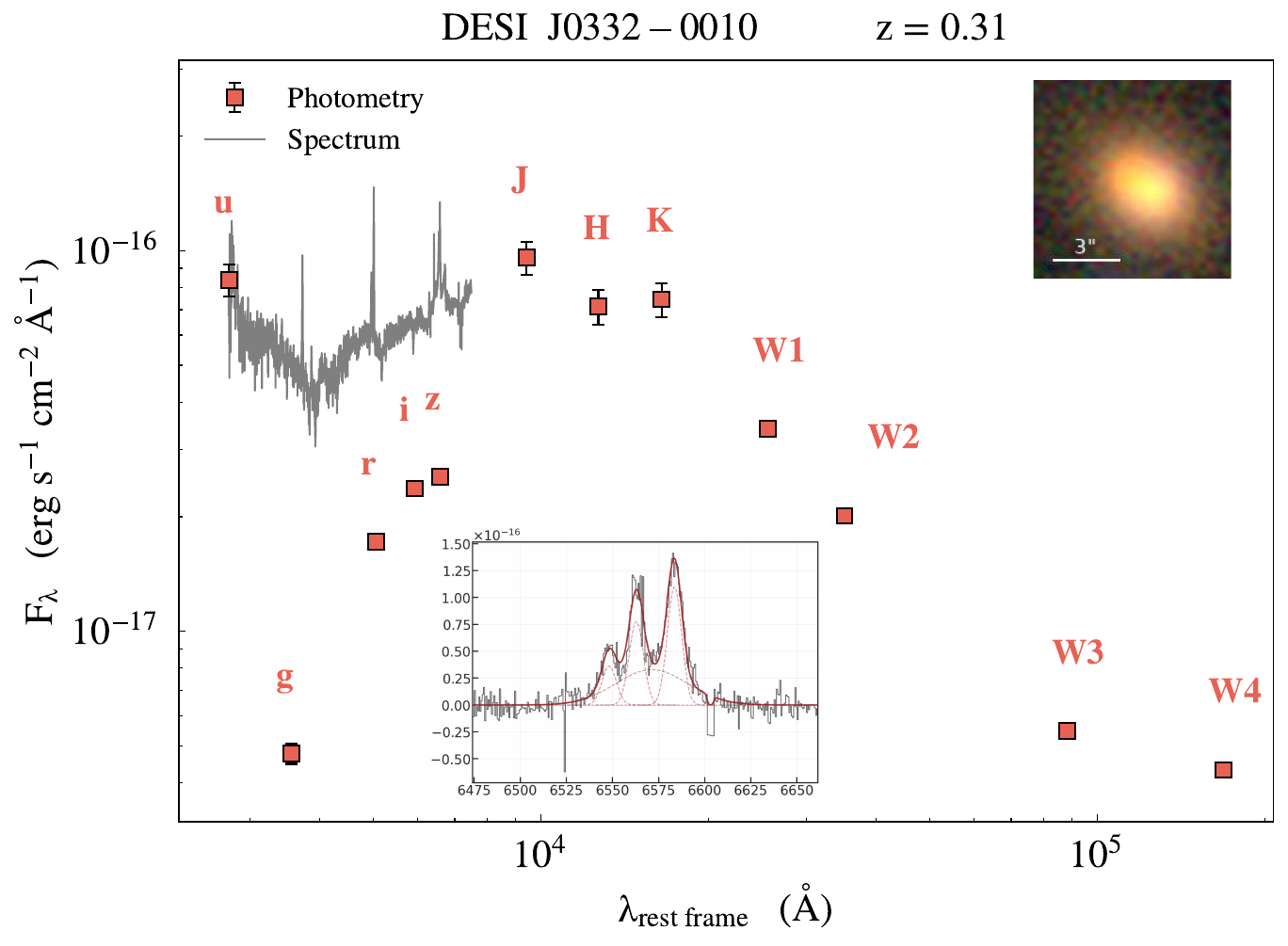}
        \includegraphics[width=0.45\linewidth]{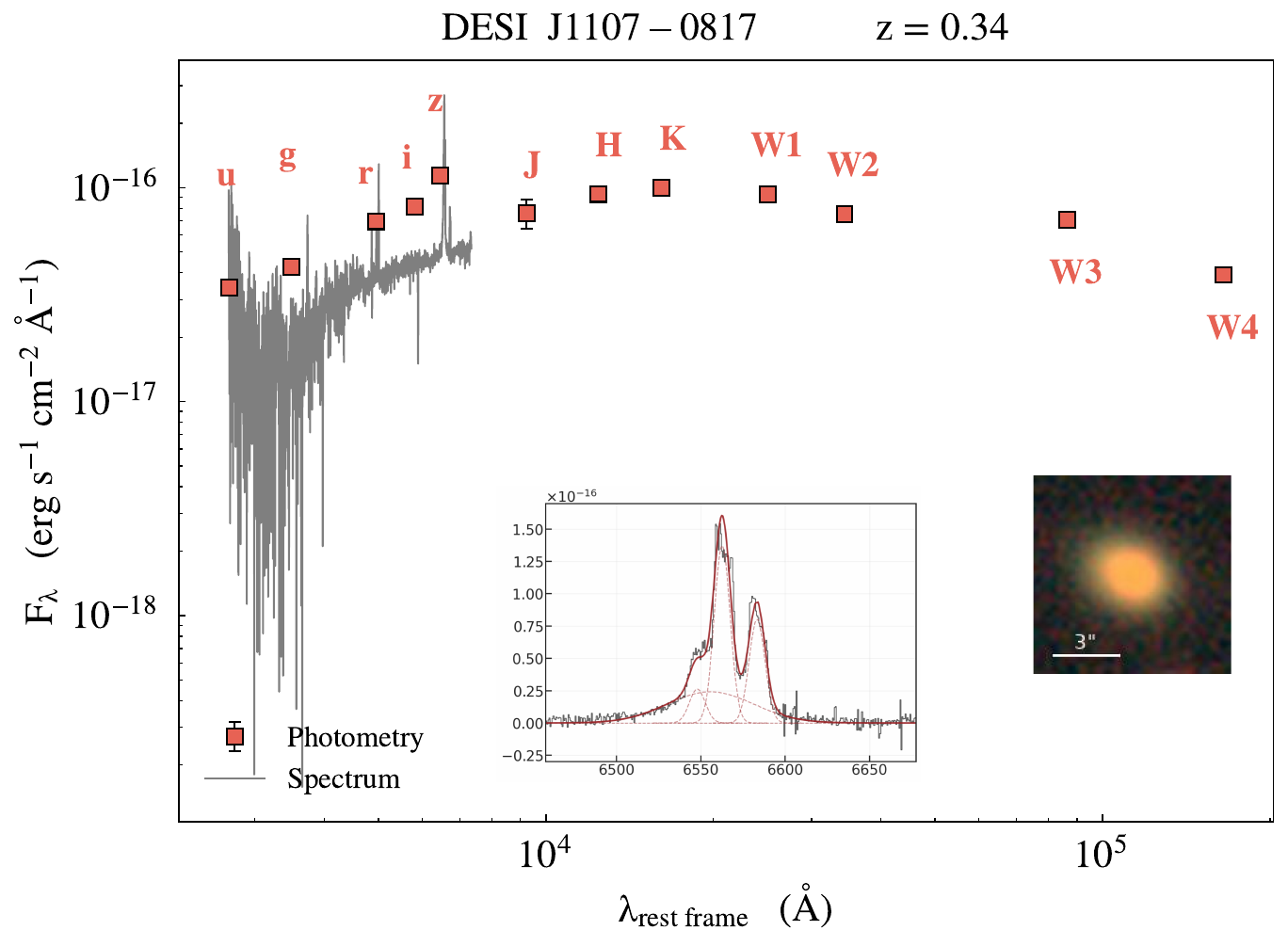}
        \caption{The images, multi-wavelength photometry, and DESI spectra of the two candidates that we excluded from the final sample.}
        \label{fig:sed_figures_others}
    \end{figure*}

Fig. \ref{fig:sed_figures} shows the multi-wavelength view of the four objects in our final sample, complementing the representative object shown in Fig. \ref{fig:sed}. The pronounced V-shaped continuum and the declining mid-infrared slope are seen in this low-redshift example.

Fig. \ref{fig:sed_figures_others} shows the two candidates that were excluded during our final selection process. The spectrum of DESI J0332$-$0010 is inconsistent with its photometric data, possibly due to aperture effects and contamination from a nearby source. For DESI J1107$-$0817, closer inspection revealed that its infrared photometry lacked the characteristic downturn seen in the other LRDs. We therefore excluded both objects to ensure the purity of our final sample.

In Fig.~\ref{fig:sed}, Fig.~\ref{fig:sed_figures}, and Fig.~\ref{fig:sed_figures_others}, all $u$-band data points represent synthetic photometry, while all other photometric points correspond to real broadband imaging measurements. We note that for DESI J2146+1028 and DESI J0332+0010, the DESI-based synthetic photometry is notably brighter than the SDSS catalog $u$-band magnitudes, and the $u$-band turnover is seen only in the DESI spectra. The differences between the DESI-based synthetic photometry and SDSS catalog magnitudes may arise from aperture effects. In particular, aperture effects may play an important role, as DESI spectra are obtained within a fixed fiber aperture, whereas SDSS model magnitudes are derived from imaging data and are intended to represent the total flux of the source. For these extremely compact red sources, the UV excess may be more efficiently captured by the fiber than by the automated imaging pipelines. Certainly, the relatively large error in the observed u-band magnitude is due to the source's faint flux in the SDSS image for that filter. This may introduce additional uncertainties and contribute to the observed differences in the $u$-band.

\section{CIGALE SED fitting}\label{sec:appendixB}

\begin{table*}
\caption{Extinction-corrected photometry of our sample.}
\label{tab:photometry}
\centering
\begin{tabular}{lccccc}
\hline\hline
  & DESI J2146$+$1028 & DESI J2357$-$1128 & DESI J1051$-$0626 & DESI J0852$-$0411 & DESI J0929$-$0251  \\
\hline
$u_{\rm syn}$ & 19.21 $\pm$ 0.09  & 19.55 $\pm$ 0.08 & 20.33 $\pm$ 0.10 & 20.59 $\pm$ 0.08 & 20.37 $\pm$ 0.09  \\
$\rm{SDSS}$   $u$ & 20.79 $\pm$ 0.11&  &  &  & 20.45 $\pm$ 0.09 \\
$\rm{DESI\  LS}$  $g$ & 19.12 $\pm$ 0.03 & 20.46 $\pm$ 0.01 & 20.50 $\pm$ 0.01 & 20.36 $\pm$ 0.08 & 20.42 $\pm$ 0.02  \\
$\rm{DESI\  LS}$  $r$ & 18.12 $\pm$ 0.02 & 18.26 $\pm$ 0.06 & 19.20 $\pm$ 0.04 & 19.76 $\pm$ 0.09 & 19.41 $\pm$ 0.03  \\
$\rm{DESI\  LS}$  $i$ & 17.90 $\pm$ 0.01 & 18.79 $\pm$ 0.03 & 18.99 $\pm$ 0.05 & 19.48 $\pm$ 0.06 & 18.96 $\pm$ 0.04  \\
$\rm{DESI\  LS}$  $z$ & 17.38 $\pm$ 0.02 & 18.46 $\pm$ 0.03 & 18.45 $\pm$ 0.04 & 19.06 $\pm$ 0.06 & 17.91 $\pm$ 0.02  \\
$\rm{2MASS}$  $\rm{J}$ & 17.19 $\pm$ 0.10 & 17.87 $\pm$ 0.17 & & 17.54 $\pm$ 0.16 & 17.74 $\pm$ 0.18  \\
$\rm{2MASS}$  $\rm{H}$ & 17.01 $\pm$ 0.11 & 17.45 $\pm$ 0.15 & & 17.68 $\pm$ 0.20 & 17.12 $\pm$ 0.15  \\
$\rm{2MASS}$  $\rm{K}$ & 18.14 $\pm$ 0.13 & 18.81 $\pm$ 0.24 & & 18.48 $\pm$ 0.11 & 18.69 $\pm$ 0.09  \\
$\rm{WISE}$  $\rm{W1}$ & 16.63 $\pm$ 0.03 & 17.64 $\pm$ 0.04 & 17.80 $\pm$ 0.04 & 17.49 $\pm$ 0.03 & 15.26 $\pm$ 0.02  \\
$\rm{WISE}$  $\rm{W2}$ & 16.47 $\pm$ 0.03 & 17.59 $\pm$ 0.04 & 18.08 $\pm$ 0.04 & 17.64 $\pm$ 0.03 & 14.80 $\pm$ 0.02  \\
$\rm{WISE}$  $\rm{W3}$ & 14.77 $\pm$ 0.03 & 16.00 $\pm$ 0.04 & 16.26 $\pm$ 0.04 & 17.23 $\pm$ 0.03 & 13.96 $\pm$ 0.02  \\
$\rm{WISE}$  $\rm{W4}$ & 13.65 $\pm$ 0.03 & 14.78 $\pm$ 0.04 & 14.43 $\pm$ 0.04 & 15.06 $\pm$ 0.03 & 13.08 $\pm$ 0.02  \\

\hline
\end{tabular}
\tablefoot{
All magnitudes are given in the AB system and have been corrected for Galactic extinction. The synthetic $u$-band magnitudes are derived from the DESI spectra and are shown for comparison. The $u$-band photometry is taken from the SDSS, the $griz$ bands from the DESI Legacy Imaging Surveys DR10, the near-infrared $JHK$ bands from 2MASS, and the mid-infrared $W1$--$W4$ bands from WISE. 
}
\end{table*}

\begin{figure*}[htbp!]
        \centering
        \includegraphics[width=0.3\linewidth]{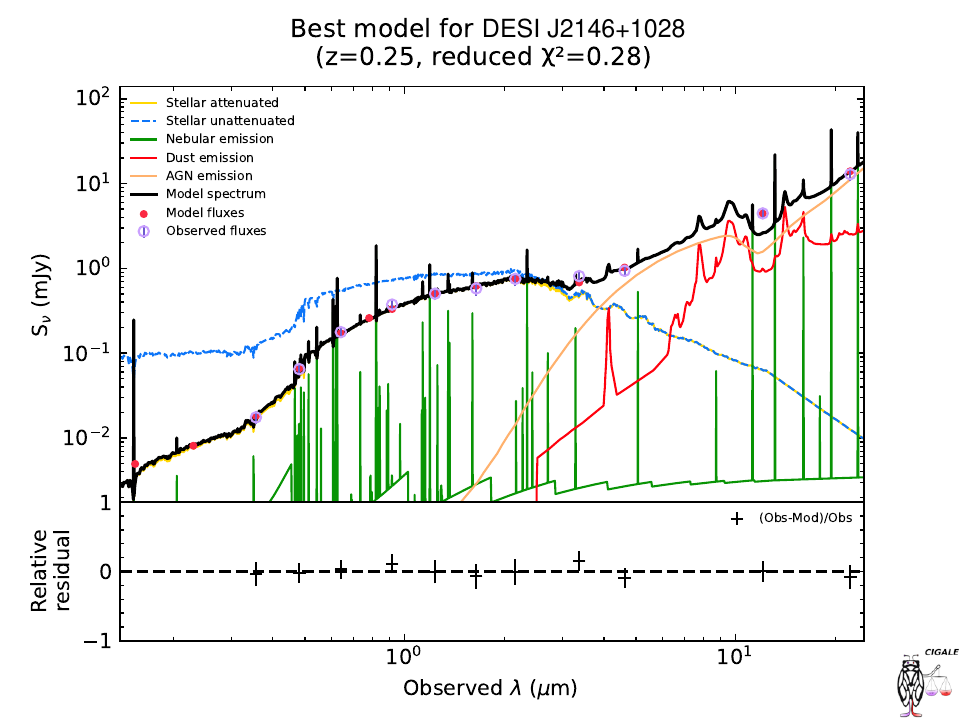}
        \includegraphics[width=0.3\linewidth]{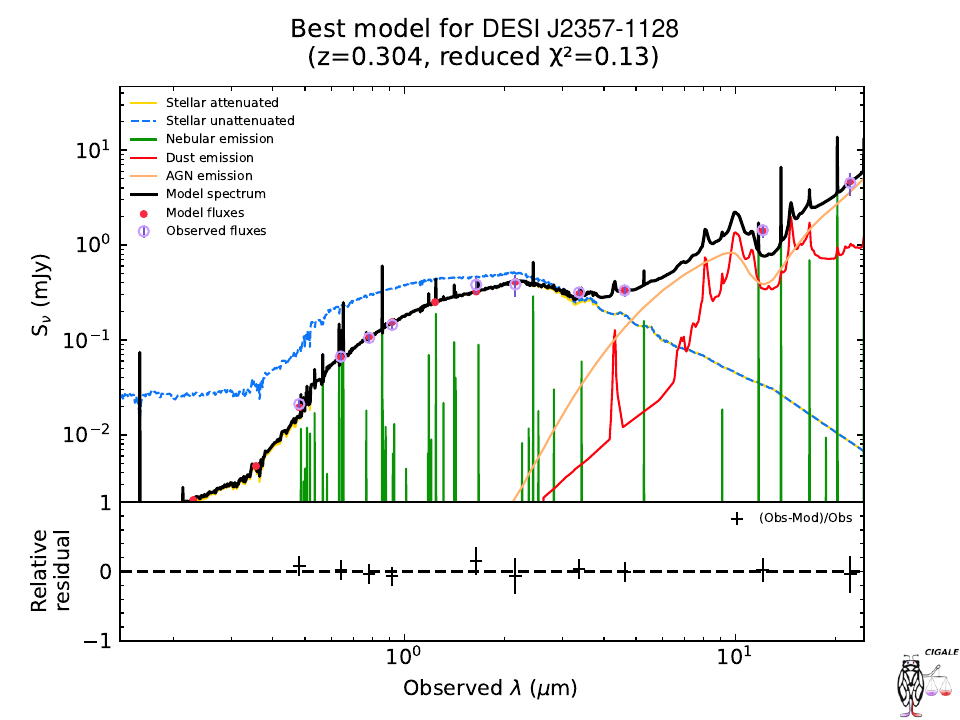}
        \includegraphics[width=0.3\linewidth]{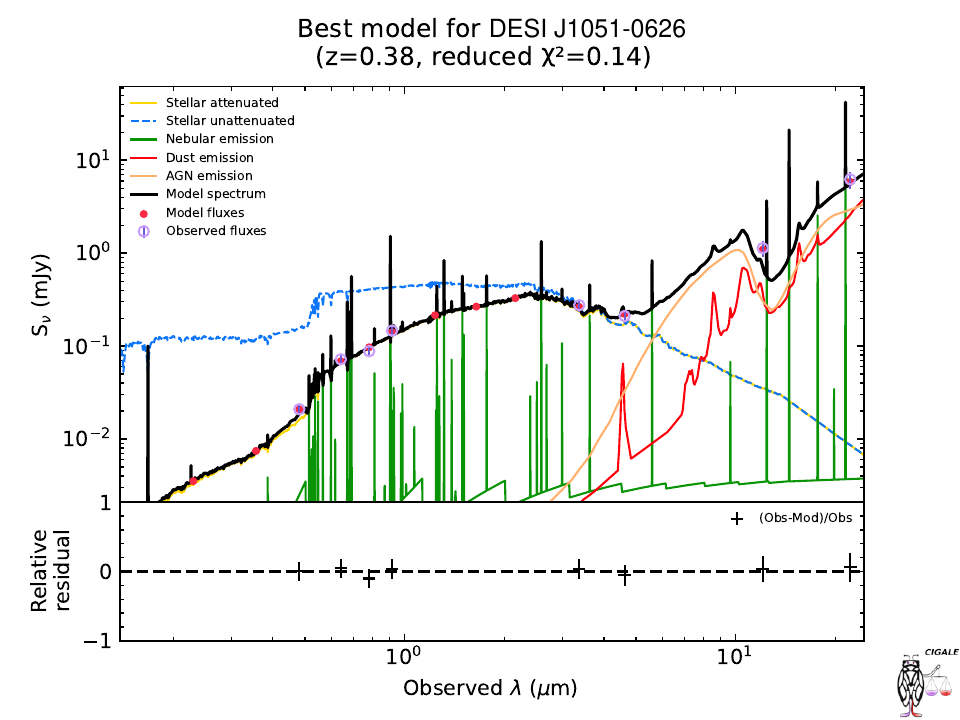}
        \includegraphics[width=0.3\linewidth]{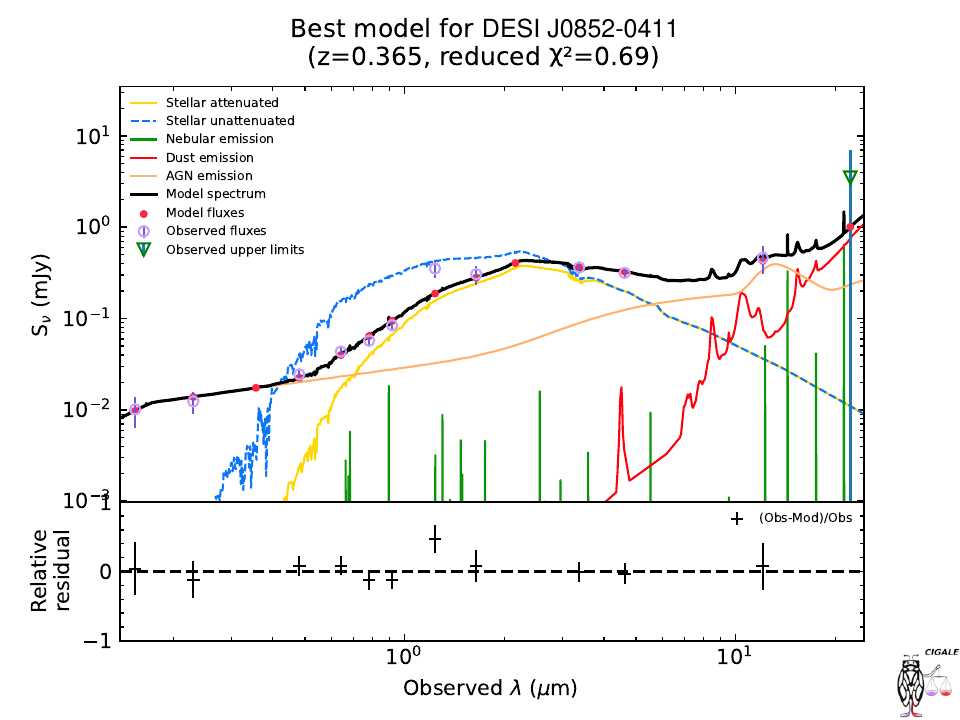}
        \includegraphics[width=0.3\linewidth]{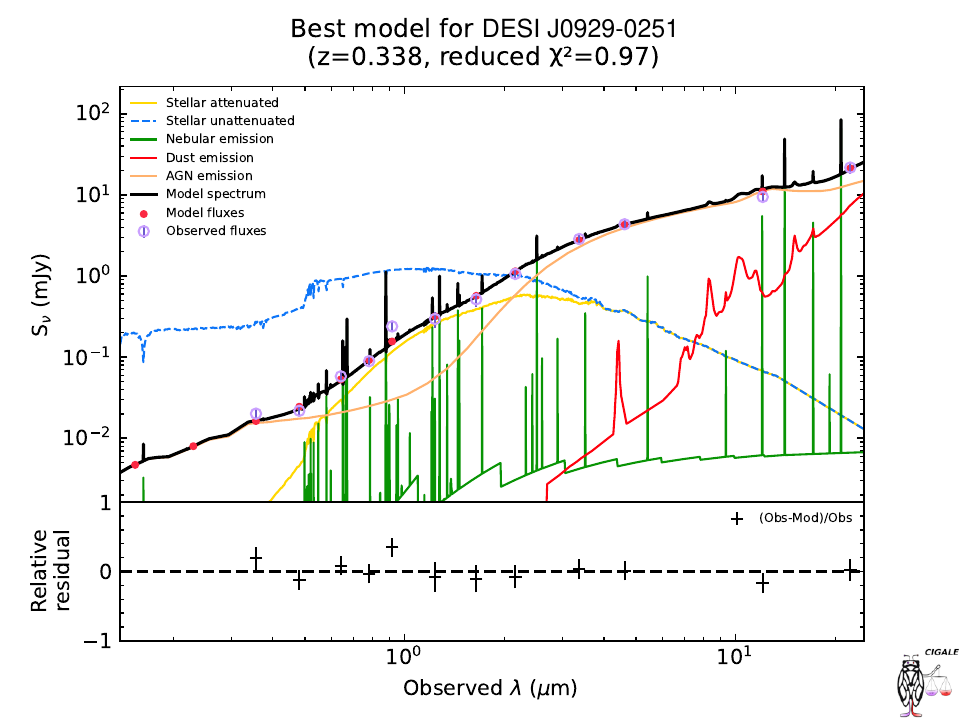}
        \caption{Multi-wavelength SED fits for the LRD analogue sample performed with \texttt{CIGALE}. The observed photometry includes SDSS $u$, DESI Legacy Imaging Surveys $griz$, 2MASS $JHK$, and WISE $W1$--$W4$ bands. In each panel, the upper panel shows the models and photometric data and the bottom panel presents the residual between the observed and model data.}
        \label{fig:cigale}
    \end{figure*}

We employ version 2022.1 of the Code Investigating GALaxy Emission \citep[CIGALE;][]{2019A&A...622A.103B, 2020MNRAS.491..740Y, 2022ApJ...927..192Y}, which reconstructs galaxy SEDs over a broad wavelength range. We emphasize that the SED fitting is based on real broadband photometry. We assume a delayed-$\tau$ star formation history with an additional burst component to model the young stellar population, the BC03 stellar population models \citep{2003MNRAS.344.1000B}, and a Chabrier initial mass function \citep{2003PASP..115..763C}. The modeling accounts for nebular and dust emission \citep{2014ApJ...780..172D}, adopting the \cite{2000ApJ...533..682C} attenuation law and the AGN templates from \cite{2016MNRAS.458.2288S}. The photometric data used are presented in Table~\ref{tab:photometry}, and the input parameter space is summarized in Table~\ref{tab:cigale}. The \texttt{CIGALE} fits are shown in Fig. \ref{fig:cigale}, providing a robust description of the observed photometry with reduced $\chi^2 < 1$ for all objects. Our results indicate that the young stellar population accounts for $<1\%$ of the total stellar mass in the entire sample, implying that the SED is not dominated by recent starburst activity.

The resulting SED decompositions for our sample exhibit qualitative differences from the "standard" interpretative framework typically adopted for high-redshift ($z \gtrsim 4$) LRDs. In that standard model, the UV emission is typically attributed to a stellar population, while the optical-to-near-infrared continuum is attributed to the AGN component \citep[e.g.,][]{2024ApJ...964...39G, 2025ApJ...978...92L}. For three of our sources (DESI J2146+1028, J2357-1128, and J1051-0626), the best-fit solutions correspond to heavily obscured AGNs. In these cases, our fits favor a configuration where a massive stellar population dominates the UV–optical continuum, while the obscured AGN emission is primarily confined to the infrared. Consequently, the host galaxy rather than the AGN is the primary contributor to the optical continuum, even in the presence of broad Balmer emission lines. Given that the underlying physical mechanism of the LRD population remains a subject of intense debate with numerous competing models, several interpretations exist that are consistent with these results. Stellar-dominated optical SEDs align with observations of certain high-redshift LRDs where the AGN does not dominate the optical bands \citep[e.g.,][]{2025ApJ...978...92L, 2025ApJ...985..119Z}, and similar fits have been achieved for systems with prominent broad H$\alpha$ lines \citep[e.g.,][]{2024ApJ...969L..13W, 2025ApJ...981..191M}. Specifically, these three sources may correspond to the model proposed by \cite{2024ApJ...968....4P}, where very efficient dust production and star formation dominate the global UV–optical output. Alternatively, they may represent an evolutionary stage where the dense gas envelope enshrouding a super-Eddington accreting SMBH has begun to disperse into a clumpy, porous structure \citep{2025arXiv251202096F}. In this scenario, substantial host-galaxy mass assembly over cosmic time, combined with the emergence of a matured dust torus, results in the observed SED configuration. Similar configurations have also been reported for other low-redshift LRD analogues \citep[e.g.,][]{2024MNRAS.535..853J, 2026MNRAS.545f2235J, 2026ApJ..1000L..18H}.    

\begin{table*}[h!] 
\caption{CIGALE parameter settings of our sample. Unlisted parameters are set to the default values.}
\label{tab:cigale}
\centering
\small
\begin{tabular}{l c c p{8cm}} 
\hline
\hline
Module & Parameter & Name in the CIGALE & Possible values\\
(1) & (2) & (3) & (4)\\
\hline
Delayed SFH & Stellar \textit{e-}folding time & tau\_main & 0.1, 0.2, 0.3, 0.4, 0.5, 0.6, 0.7, 0.8, 0.9, 1, 2, 3, 4, 5, 6, 7, 8, 9, 10 Gyr \\
& Stellar age & age\_main & 0.1, 0.2, 0.3, 0.4, 0.5, 0.6, 0.7, 0.8, 0.9, 1, 2, 3, 4, 5, 6, 7, 8, 9, 10 Gyr \\
& Burst \textit{e-}folding time & tau\_main & 10, 30, 50 Myr \\
& Burst age & age\_burst & 10, 30, 50 Myr \\
&Mass fraction of burst population & f\_burst & 0.0, 0.1, 0.5, 0.9 \\
\hline
Simple stellar & Initial mass function & imf & 1\\
population & Metallicity & metallicity & 0.0001, 0.0004, 0.004, 0.008, 0.02, 0.05\\
\hline
Nebular & ... & ... & ... \\
\hline
Dust attenuation & $E(B-V)_\mathrm{line}$ & E\_BV\_lines & 0.001, 0.01, 0.1, 0.2, 0.3, 0.4, 0.5, 0.6, 0.8, 1, 1.2 \\
& $E(B-V)_\mathrm{line}/$ & E\_BV\_factor & 1 \\
& $E(B-V)_\mathrm{continuum}$ & & \\ 
\hline
Dust emission & Alpha slope & alpha & 1.0, 2.0, 3.0 \\
\hline
AGN & Edge-on optical depth & t & 3, 7, 11 \\
& Viewing angle & i & $0^\circ, 10^\circ, 30^\circ, 50^\circ, 70^\circ, 90^\circ$ \\
& Disk spectrum & disk\_type & 1 \\
& Mod. of opt. power-law & delta & $-0.27$ \\ 
& AGN fraction & fracAGN & 0.1, 0.2, 0.4, 0.6, 0.8, 0.99 \\
& $E(B-V)$ polar ext. & EBV & 0, 0.05, 0.1, 0.2, 0.3, 0.4, 0.5\\
\hline
\end{tabular}
\end{table*}

For the remaining two sources, DESI J0852-0411 and J0929-0259, the SEDs are best described by an unobscured AGN dominating the UV and IR emission, while a massive stellar population provides the primary contribution to the optical continuum. Although this configuration deviates from the "standard" high-redshift LRD paradigm, literature such as \cite{2025ApJ...995..118S} demonstrates that unobscured AGN models can provide statistically sound fits to LRD-like signatures. Similar SED fitting results have also been reported in other studies of local LRD analogues \citep[e.g.,][]{2026MNRAS.545f2235J}, suggesting that such component allocations are not isolated to this sample. An alternative interpretation is that these sources represent a late evolutionary stage of LRDs that originally hosted super-Eddington accreting SMBHs enshrouded in dense and dust-poor gas. In this scenario, the gaseous cocoon has nearly dissipated, allowing the previously obscured AGN emission to emerge in the UV \citep{2025arXiv251202096F}. Concurrently, substantial host-galaxy mass assembly over cosmic time results in a dominant stellar component in the optical. This evolutionary transition leads to an SED that resembles a typical Type-1 AGN while still retaining the distinctive features of the LRD population. Alternatively, the diversity in the resulting SED decompositions may indicate that these objects arise from physical mechanisms entirely distinct from those of the high-redshift LRD population. A possible contamination from a foreground galactic star may affect the observed SED shape. As shown in Fig. \ref{fig:sed_figures}, mild contamination cannot be entirely excluded for DESI J0852$-$0411. We therefore caution that the SED shape, particularly in the optical bands, may be subject to additional uncertainties.

Overall, the diversity of SED fitting solutions within our sample highlights the inherent degeneracies and complexities involved in interpreting compact red sources. As discussed in Section \ref{subsec:discussion}, these objects do not necessarily represent direct, one-to-one local analogues of the high-redshift LRD population. Instead, they may be systems that require SED interpretations different from those typically adopted for the high-redshift population. While the diverse configurations could reflect different evolutionary stages of a common progenitor (specifically the transition from a heavily enshrouded phase to a more mature, "clearing" AGN phase) this interpretation remains one of several viable possibilities. We note that reliably decomposing the host-galaxy stellar component from unresolved broadband photometry is a recognized challenge, particularly for the unobscured AGN cases where the nuclear emission can significantly contaminate the optical continuum. This introduces inherent uncertainties into the inferred stellar mass estimates. However, the inclusion of \textit{WISE} mid-infrared photometry provides critical constraints on the dust emission and AGN torus, which helps mitigating the degeneracies between the AGN and stellar components. Sensitivity tests using various parameter sets indicate that while the stellar mass estimates are model-dependent, the variations remain within approximately 1 dex (consistent with other low-redshift studies, \citep[e.g.][]{2026MNRAS.545f2235J}), and do not alter the fundamental conclusions regarding the host-galaxy properties in Section \ref{subsec:mass}. Consequently, this study aims to document and contextualize the observational diversity of LRD-like signatures at low redshift, acknowledging that multiple physical mechanisms may contribute to the observed population.

\section{Properties of our sample}\label{sec:appendixC}

\begin{table}[h!]
\caption {Properties of our sample}
\label{table:properties} 
\centering
\begin{tabular}{ccccccccc}
\hline\hline    
\\
DESI ID & RA & Dec. & Redshift &  R$_{50}$ & log($L_{\rm{H\alpha,\,  broad}}$) &log$(M_*/M_{\odot})$ &  log$(M_{\rm{BH}}/M_{\odot})$ & 12+log(O/H)\\
\\
  & (J2000)& (J2000) &  & (kpc) & ($\mathrm{erg \, s^{-1}}$)& &  &\\
  \\
\hline
\\
        DESI J2146$+$1028 & 326.73214 & 10.47036 & 0.25 & 3.53  & 43.43 & 10.76 $^{+0.14}_{-0.14}$ & 7.86 $\pm$ 0.09 & 8.94 $\pm$ 0.22 \\ 
        \\
        DESI J2357$-$1128 & 359.46573 & -11.48030 & 0.30 & 3.14  & 42.70 & 10.93 $^{+0.17}_{-0.16}$ & 6.85 $\pm$ 0.14 & 8.78 $\pm$ 0.19\\
        \\
        DESI J1051$-$0626 & 162.99046 & -6.44040 & 0.38 & 2.71  & 43.81 & 10.74 $^{+0.27}_{-0.26}$ & 8.02 $\pm$ 0.15 & 8.95 $\pm$ 0.24\\ 
        \\
        DESI J0852$-$0411 & 133.18388 & -4.19739 & 0.37 & 3.07 & 43.71 & 11.18 $^{+0.12}_{-0.12}$ & 7.97 $\pm$ 0.14 & 8.74 $\pm$ 0.20 \\ 
        \\
        DESI J0929$-$0251 & 142.27076 & -2.86616 & 0.34 & 0.97 & 45.10 &  11.16 $^{+0.12}_{-0.15}$ & 8.96 $\pm$ 0.25 & 8.63 $\pm$ 0.18\\ 
        \\

\hline
\end{tabular}
\tablefoot{(1) Object name;  (2-3) right ascension and declination in J2000 from DESI DR1; (4) DESI spectroscopic redshift; (5) Half-light radius of galaxy model; (6) Logarithmic stellar mass in $M_{\odot}$; (7) Logarithmic black hole mass in $M_{\odot}$; (8) Metallicity estimated with the [N {\sc ii}] line}
\end{table}

\end{appendix}
\end{document}